\documentclass[twocolumn,aps,prc,tightenlines,floats,floatfix,nofootinbib]{revtex4}

\def\bea{\begin{eqnarray}}
\def\eea{\end{eqnarray}}

\def\etal{{\it et al.\/}$\;$}
\def\sfrac#1#2{{\textstyle \frac{#1}{#2}}}

\def\be{\begin{equation}}
\def\ee{\end{equation}}
\def\ba{\begin{eqnarray}}
\def\ea{\end{eqnarray}}

\usepackage{graphics}
\usepackage{graphicx}
\usepackage{epsf}
\usepackage{amsmath}
\usepackage{amssymb}

\begin{document}

\phantom{0}
\vspace{-0.2in}
\hspace{5.5in}

\vspace{-1in}

\title
{\bf 
Extracting the $\Omega^-$ electric quadrupole moment from lattice QCD data}
\author{G. Ramalho$^{1}$ and
M.~T.~Pe\~na$^{1,2}$
\vspace{-0.1in}  }

\affiliation{
$^1$CFTP, Instituto Superior T\'ecnico, 
Av.\ Rovisco Pais, 1049-001 Lisboa, Portugal \vspace{-0.15in}}
\affiliation{
$^2$Department of Physics, Instituto Superior T\'ecnico, 
Av.\ Rovisco Pais, 1049-001 Lisboa, Portugal}

\vspace{0.2in}
\date{\today}

\phantom{0}

\begin{abstract}
The $\Omega^-$ has an extremely long lifetime, and is the most stable 
of the baryons with spin 3/2.
Therefore the $\Omega^-$ magnetic moment is very accurately known.
Nevertheless, its electric quadrupole moment was never measured,
although estimates exist in different formalisms.
In principle,  lattice QCD simulations provide 
at present the most appropriate way to estimate 
the $\Omega^-$ form factors,
as function of  the square of the transferred four-momentum, $Q^2$,  
since it describes
baryon systems at the physical mass for the strange quark.
However, lattice QCD form factors, and
in particular $G_{E2}$, are determined at 
finite $Q^2$ only, and 
the extraction of the electric quadrupole moment,
$Q_{\Omega^-}= G_{E2}(0)\sfrac{e}{2 M_\Omega}$, 
involves an extrapolation of the numerical lattice results.
In this work we reproduce the lattice QCD data 
with a covariant spectator quark model for $\Omega^-$ which
includes a mixture of S and 
two D states for the relative quark-diquark motion.
Once the model is calibrated, it is used
to determine $Q_{\Omega^-}$. 
Our prediction is $Q_{\Omega^-}= (0.96\pm0.02)\times 10^{-2}$ $e$fm$^2$
[$G_{E2}(0)=0.680\pm0.012$]. 
\end{abstract}

\vspace*{0.9in}  
\maketitle

\section{Introduction}

The prediction of the electromagnetic structure of
baryons and mesons is an important challenge for quark models and, 
when compared to
the available experimental results, it provides a
test on the relevant hadronic degrees of freedom.
For baryons with spin 1/2 (as in the baryon octet)
the charge and the magnetic
moment are the only multipole moments to be defined, while
for  baryons with spin 3/2 (as in the baryon decuplet)
also the electric quadrupole moment exists.
However, at present, there is no experimental measurement 
of the electric quadrupole moment for any of the baryons,
although there are several model predictions 
for the $\Delta$ and the  $\Omega^-$,  and other decuplet particles.
For a summary of the $\Delta$ results see 
Refs.~\cite{DeltaDFF,DeltaDFF2}.
As for the $\Omega^-$ there are predictions 
based in quark models 
\cite{Gershtein81,Richard82,Isgur82,Leonard90,Krivoruchenko91,Wagner00,Dahiya10},
chiral perturbation theory \cite{Butler94,Geng09},
large-$N_c$ limit \cite{Buchmann02,Buchmann03,Buchmann07} 
and other formalisms 
\cite{Kroll94,Oh95,Aliev09,Nicmorus10}.

Within all members of the baryon decuplet, 
the $\Omega^-$ is specially interesting.
As it is composed solely by strange quarks
in the valence sector, 
it can decay only by weak interaction and therefore its 
lifetime is extremely longer than
the one of the other baryons.
For this reason, experimentally, the $\Omega^-$  properties
are easier to be determined than the ones of any other member of the decuplet.
A good illustration of this is the accuracy of the 
$\Omega^-$ magnetic moment 
$\mu_{\Omega^-}=-(2.019\pm 0.053)\mu_N$  
\cite{PDG,Diehl91,Wallace95,Omega},
where $\mu_N$ is the nuclear magneton.
It is also expected that the $Q_{\Omega^-}$ 
quadrupole moment will be measured in a near future
\cite{Sternheimer73,Baryshevsky93,Karl07,Krivoruchenko08}.

Additionally, the valence quark content of the $\Omega^-$
is restricted to strange quarks, with a mass considerably 
larger than the light $u$ and $d$ quark masses.
As now
it is already possible to perform lattice
QCD at the physical strange quark mass,  the $\Omega^-$ 
magnetic moment \cite{Bernard82,Leinweber92,Aubin},
and more recently also
its electric charge $G_{E0}$, magnetic dipole $G_{M1}$,
electric quadrupole $G_{E2}$ and magnetic octupole 
$G_{M3}$ \cite{Alexandrou10,Boinepalli09} 
were calculated within lattice QCD.
Another important issue is that
in sea quark effects for the $\Omega^-$ only at most one single 
light quark participates, and therefore
the pion has no role in this case.
As in chiral perturbation theory loops involving mesons heavier 
than the pion are suppressed,  
the $\Omega^-$ becomes then a special case where 
meson cloud corrections to the valence quark core
are expected to be small.
A consequence of the smallness of the meson cloud 
effects is that lattice QCD simulations, 
quenched or unquenched, should be a good approximation
to $\Omega^-$ form factors at the physical point.
Therefore in this work we take the lattice QCD simulations
as good representations of the physical results,
without any extrapolation of the lattice data to the physical pion mass.

The main limitation in obtaining the $\Omega^-$ 
electromagnetic form factors in lattice QCD 
simulations comes from these ones being restricted, for practical reasons, to
finite non-zero values of $Q^2$, while
the determination of the quadrupole moment, for instance, 
is proportional to  $G_{E2}(0)$.
An extrapolation in the momentum transfer squared $Q^2$, 
down to $Q^2=0$ is then required, 
and one has inevitably to resort to an analytical form to do it. 
It is at this point that it is reasonable to expect that
a quark model is useful, in particular for systems
as the $\Omega^-$, without light valence quarks and where
meson loop corrections are expected to be small.
Since the covariant spectator quark model was
tested already for spin 1/2 baryons \cite{Nucleon,FixedAxis,Octet,Lattice}, 
spin 3/2 baryons 
\cite{DeltaDFF,DeltaDFF2,Omega,DeltaFF0,Delta1600},  
including strange quarks, 
and electromagnetic transitions between different 
baryon states 
\cite{NDelta,NDeltaD,Lattice,LatticeD,Roper,Delta1600,ExclusiveR},
it is a good candidate to be used 
not only to interpolate between lattice QCD data in $Q^2$,
but also to extrapolate the form factor data to $Q^2=0$.

The covariant spectator quark model was applied in the past 
to estimate successfully the leading order form factors 
of the $\Omega^-$ ($G_{E0}$ and $G_{M1}$) neglecting  
D-state admixtures  \cite{Omega}.
Here, we extend the formalism to the case where 
the D-state admixture coefficients are non-zero. 
Because these states are included, 
we have now contributions for the 
$G_{E2}$ and $G_{M3}$ form factors,
and therefore we can use the model to extract 
the $\Omega^-$ electric quadrupole moment from lattice QCD data.
By adjusting some parameters associated 
with the $\Omega^-$ wavefunction to the lattice data,
our procedure has the advantage of incorporating into the 
phenomenological model the 
fundamental theory of the strong interaction, 
in its discrete version (lattice QCD). 
The information on the 
wavefunction parameters
allows us then to calculate  all the 
electromagnetic form factors $G_{E0}$, $G_{M1}$,
$G_{E2}$ and $G_{M3}$ as functions of $Q^2$.
In particular the model can be used to 
determine the electric quadrupole 
form factor at the  $Q^2=0$ point.
To constrain the model we use the unquenched lattice QCD data 
from Ref.~\cite{Alexandrou10} for $\Omega^-$ 
at the physical $\Omega^-$ mass.
Although the existing lattice data is unquenched,
the valence quarks are expected to play the main role 
and meson dressing to be small, as mentioned before.
This is why the adjustment of the quark model 
to the lattice QCD data is meaningful.
To better constrain the parameters of the quark model, 
we use also the single datapoint for $G_{M3}$
from Ref.~\cite{Boinepalli09},  
in addition to the lattice data from Ref.~\cite{Alexandrou10}.

We start by using the spectator formalism to 
represent the $\Omega^-$ wavefunction,  
in a  similar way  to the one used before for the $\Delta$
\cite{NDeltaD}. Although both the $\Delta$ and 
the $\Omega^-$ have the same spin structure, they differ in 
their flavor content.  Therefore
we begin with the SU(3) generalization of the spectator quark model  
for the overall study of the  baryon decuplet  \cite{Omega}. 
Also, the quark momentum distributions for the 
two baryons are different, and our calculation at the 
end quantifies this difference.
As for the 
electromagnetic current associated with 
the interaction of the photon with the strange quark, 
which is another aspect where the
calculation differs from the calculation for the $\Delta$ baryon,  
we use the current based on vector meson dominance from Ref.~\cite{Omega}.

Under the assumption that the D-state components are small, 
then we take only the electromagnetic current
matrix elements which are in first order in the admixture coefficients
as done in Ref.~\cite{DeltaDFF2} for the $\Delta$.
Finally, in the process of adjusting the $\Omega^-$ electromagnetic 
form factor results to the lattice data, we determine
the percentage of each D state present in the orbital 
quark-diquark part of the wavefunction of the $\Omega^-$ baryon.
Our calculation enables us at the end to narrow the 
uncertainty in the extraction of the value of the quadrupole magnetic moment
of the $\Omega^-$ from the lattice QCD data.

This work is organized as follows:
In Sect.~\ref{secFormFactors} we give
the formulas for the 
$\Omega^-$ electromagnetic form factors 
in first order of the admixture coefficients;
In Sect.~\ref{secWF} we parametrize
the $\Omega^-$ wavefunction and its 
momentum dependence.
The results are presented in Sect.~\ref{secResults}
and the conclusions and final remarks 
in Sect.~\ref{secConclusions}.

\section{$\Omega^-$ form factors}
\label{secFormFactors}

We use here the covariant spectator quark model, where
relativity is implemented consistently.
Within this framework a baryon is described 
described as a off-shell quark 
and two {\it noninteracting} on-shell spectator quarks.
Integrating over the on-mass-shell quarks 
degrees of freedom one represents those quarks states 
as a single on-shell particle (or diquark) 
with an average mass $m_D$ \cite{Nucleon,Omega}.
With this reduction, 
the wavefunction associated with the baryon states
including the spin, angular momentum, 
coordinate space and flavor structure,
can then be written as the direct product 
of the diquark and quark states properly symmetrized.

The electromagnetic interaction with the $\Omega^-$ is,
in relativistic impulse approximation,
written as the sum over the terms in which 
the photon couples to each (off-shell) quark 
in turn with the other two (on-shell) quarks 
that compose the diquark.
The electromagnetic structure of the quarks 
is parametrized in terms of form factors. 
One can re-arrange the contributions to the 
electromagnetic current from the conveniently
symmetrized wavefunctions in terms of the 
on-shell diquark states -- quark pair (12), 
and  the off-shell quark  -- quark 3 states.
The final result for the current  becomes then three times 
the current associated to the interaction 
with quark 3. 
See Ref.~\cite{Omega} for a detailed discussion.
We write then the $\Omega^-$ wavefunction 
as $\Psi_\Omega(P,k)$  for total momentum $P$,  
as the combination 
of the diquark (on-shell) states, with momentum $k$,
and the quark 3 (off-shell) states.  
%
In our notation  the indices for the diquark polarization
$\epsilon =0,\pm$ and the $\Omega^-$ spin projection
are omitted for simplicity, and
the matrix element of the electromagnetic current
between the  initial and final states of
momentum $P_+$  and $P_-$, respectively, is written as
\be
J^\mu =
3 \sum_\epsilon \int_k \overline \Psi_\Omega (P_+,k) 
j_q^\mu \Psi_\Omega(P_-,k),
\label{eqJ0}
\ee
where $j_q^\mu$ is the current operator for quark 3
and  $\int_k$ is the covariant integral is defined as
$\int_k = \int \sfrac{d^3 k}{(2\pi)^2 2 E_D}$, 
where $E_D$ is the diquark on-shell energy. 
In Eq.~(\ref{eqJ0}) the interactions with {\it all}
 quarks are counted, without including the coupling 
with the diquark \cite{Omega,Nucleon}.

In this work the $\Omega^-$ wavefunction 
is represented as a 
combination of an S state 
and two D states for the quark-diquark relative motion \cite{NDeltaD}:
\be
\Psi_\Omega(P,k)=
N\left[\Psi_S(P,k)+ a \Psi_{D3}(P,k) + b\Psi_{D1}(P,k) \right].
\label{eqPsiOm}
\ee
In the previous equation $a$ and $b$ are 
the mixture coefficients of the states: D3 
(core spin 3/2) and D1 (core spin 1/2) respectively,
and $N=1/\sqrt{1+a^2+b^2}$ a normalization constant.
Each of the three wavefunction terms includes a scalar wavefunction, 
respectively $\psi_S$, $\psi_{D3}$ and 
$\psi_{D1}$,  which  is a  function of 
the baryon momentum $P$ and diquark momentum $k$, with
a form and normalization given in the next section.

The quark current $j_q^\mu$ can be in general decomposed
as
\be
j_q^\mu(Q^2) = j_1 (Q^2)\gamma^\mu + 
j_2(Q^2) \frac{i \sigma^{\mu \nu}q_\nu}{2M_N},
\ee
where $M_N$ is the nucleon mass, and 
\be
j_i= \frac{1}{6}f_{i+} \lambda_0 
+  \frac{1}{2}f_{i-} \lambda_3 + 
\frac{1}{6}f_{i0} \lambda_s,
\ee
with $\lambda_0=\mbox{diag}(1,1,0)$: 
 $\lambda_3=\mbox{diag}(1,-1,0)$ and 
 $\lambda_s=\mbox{diag}(0,0,-2)$ 
are SU(3) flavor operators acting 
in the particular quark flavor state 
$q=  \left(u \, d \, s \right)^T$ \cite{Omega,Octet}.

The quark form factors are 
normalized as $f_{1 \pm}(0)=1$, $f_{1 0}(0)=1$,  
$f_{2\pm}(0)= \kappa_\pm$ and $f_{2 0}(0)= \kappa_s$.
We model the electromagnetic structure of the quarks by means of a
parametrization that is based on vector meson dominance.
In particular the strange quark form factors are 
represented \cite{Omega} by
\ba
& &
f_{10}=\lambda + (1-\lambda) \frac{m_\phi^2}{m_\phi^2+ Q^2}+
c_0 \frac{M_h^2 Q^2}{(M_h^2+ Q^2)^2}, \\
& &
f_{20}=\kappa_s 
\left\{
d_0
\frac{m_\phi^2}{m_\phi^2+ Q^2}+
(1-d_0) \frac{M_h^2}{M_h^2+ Q^2}\right\},
\ea
where $m_\phi$ is a mass of the $\phi$ meson (system $\bar s s$),
$M_h=2 M_N$ is an effective vector meson that 
simulates the short range structure and
$\lambda$ is fixed by 
deep inelastic scattering, 
and corresponds to the quark number density \cite{Nucleon}.
The coefficients defining the current were fixed as 
$c_0=4.427$ and $d_0=-1.860$ in the study 
of the dominant form factors of the baryon decuplet \cite{Omega}.  
As for the strange quark anomalous magnetic moment 
 one uses $\kappa_s=1.462$ to reproduce the experimental value of 
$\mu_{\Omega^-}$ \cite{Omega}.
The explicit expression for the  remaining
quark form factors 
are presented in Refs.~\cite{NDeltaD,LatticeD,Omega}.
 
Following Refs.~\cite{NDeltaD,LatticeD} 
the current can be written 
in terms of {\it charge} $\tilde e_\Omega$ and 
{\it anomalous magnetic moment} $\tilde \kappa_\Omega$
functions 
\ba
\tilde e_\Omega &=& 
-f_{10}(Q^2) \\
\tilde \kappa_\Omega &=& 
-f_{20}(Q^2) \sfrac{M_\Omega}{M_N}.
\ea
In the previous equation, 
for convenience we use a normalization 
that differs from the one presented in \cite{Omega}, 
by a factor $\sfrac{M_\Omega}{M_N}$.
This redefinition does not change the results.
We define also 
\ba
\tilde g_\Omega &=& \tilde e_\Omega - \tau  \tilde \kappa_\Omega \\
\tilde f_\Omega &=& \tilde e_\Omega + \tilde \kappa_\Omega,
\ea
where $\tau=Q^2/(4M_\Omega^2)$.

Working the algebra for the current 
as in Refs.~\cite{DeltaDFF,DeltaDFF2},
replacing $\tilde e_\Delta$ and $\tilde k_\Delta$
by $\tilde e_\Omega$ and $\tilde k_\Omega$,
one obtains in first order 
in the admixture coefficients $a$ and $b$:
\ba
G_{E0}(Q^2) &=&  N^2 \tilde g_\Omega {\cal I}_S  
\label{eqGE0a}  \\
G_{M1}(Q^2) &=&  
N^2 \tilde f_\Omega 
\left[
{\cal I}_S 
+\frac{4}{5} a  {\cal I}_{D3}
-\frac{2}{5} b {\cal I}_{D1} \right]
 \label{eqGM1a} \\
G_{E2}(Q^2) &=& 
N^2 \tilde g_\Omega (3a) \, \frac{{\cal I}_{D3}}{\tau} 
\label{eqGE2a} \\
G_{M3}(Q^2) &=& 
\tilde f_\Omega N^2\left[ 
a \frac{{\cal I}_{D3}}{\tau} +
2  b \frac{{\cal I}_{D1}}{\tau} \right],
\label{eqGM3a}
\ea
where the overlap between the S-states is
\be
{\cal I}_S=
\int_k \psi_S(P_+,k) \psi_S(P_-,k),
\ee
 and the overlap between the S and each one of the D-states is 
\ba
& &
{\cal I}_{D1}=
\int_k b(\tilde k_+,\tilde q_+)\psi_{D1}(P_+,k) \psi_S(P_-,k) 
\label{eqID1}
\\
& &
{\cal I}_{D3}=
\int_k b(\tilde k_+,\tilde q_+)\psi_{D3}(P_+,k) \psi_S(P_-,k).
\label{eqID3}
\ea 
The function 
$b(\tilde k_+,\tilde q_+)$ is defined 
in Ref.~\cite{NDeltaD} 
and includes the specific angular dependence of a D-state,  
with $\tilde k_+$ defined as the three-momentum
in the frame where $P_+=(M_\Omega,0,0,0)$:
$\tilde k_+ = k -\sfrac{ P_+ \cdot k}{M_\Omega^2}P_+$.

From Eq.~(\ref{eqGE0a}) one obtains that
the result for the 
charge form factor at $Q^2=0$ is $G_{E0}(0)=- N^2$,
[note that $N^2=1/(1+a^2+b^2)$],
which differs from the exact result $(-1)$, if $a,b \ne 0$.
This deviation, although small if $a$ and $b$ 
are small, 
is a consequence of taking in the calculation of the current matrix elements
only the terms in first order in 
these D state admixture coefficients $a$ and $b$.
Also  the magnetic dipole form factor $G_{M1}$, 
from Eq.~(\ref{eqGM1a}), at $Q^2=0$  
deviates slightly from the
experimental value (which gives the $\Omega^-$ magnetic moment).
%
Once the terms for the current matrix elements for the 
D- to D-state transitions are included,
the exact results of both the charge and magnetic moment  
are recovered exactly. 
We will use this fact to estimate 
the error of our model, as explained in Sect.~\ref{secResults}.

\section{Model for the scalar wavefunctions}
\label{secWF}

To describe the momentum dependence 
of the scalar wavefunctions one assumes a certain form
(as done already for the $\Delta$  in
Refs.~\cite{NDeltaD,LatticeD})
\ba
& &\psi_S(P,k)= \frac{N_S}{m_D(\alpha_S+ \chi_{_\Omega})^3}\\
& &\psi_{D3}(P,k)= \frac{N_{D3}}{m_D^3(\alpha_{D3}+ \chi_{_\Omega})^4}\\
& &\psi_{D1}(P,k)= \frac{N_{D1}}{m_D^3(\alpha_{D1}+ \chi_{_\Omega})^4},
\ea
where
\be
\chi_{_\Omega}=\frac{(M_\Omega -m_D)^2-(P-k)^2}{M_\Omega m_D}.
\ee
In this way we introduce momentum scale parameters 
($\alpha_S,\alpha_{D3}, \alpha_{D1}$)
for each angular momentum or orbital state.
In contrast to the  $\Delta$ case
one does not have to impose the orthogonality of the spin core 
$S=1/2$ D-state with an S-state with the same spin, as a spin 1/2 S-state with
three strange quarks is forbidden. Therefore,
the expression for the state D1 differs from the
one used for the $\Delta$ baryon  \cite{NDeltaD,LatticeD},
and 
one may consider a simpler parameterization for the $\Omega^-$ D1 state.

The normalization conditions are given by
\ba
& &
\int_k |\psi_S(\bar P,k)|^2=1 \\
& &
\int_k |\tilde k^2\psi_{D3}(\bar P,k)|^2=1 \\
& &
\int_k |\tilde k^2\psi_{D1}(\bar P,k)|^2=1,
\ea
where $\bar P=(M_\Omega,0,0,0)$ is the 
momentum of the $\Omega^-$ in the rest frame.
The parameters $\alpha_S,\alpha_{D3}, \alpha_{D1}$
and the admixture coefficients $a$, $b$ are 
the five adjustable parameters  of the model.

\section{Results}
\label{secResults}

In this section we present the results for the  $\Omega^-$ 
electromagnetic form factor data obtained from the quark model 
described in the previous sections, calibrated
by a fit to the available lattice data.
The available $\Omega^-$ form factor data 
as function of $Q^2$, is restricted 
to the unquenched lattice simulations of the 
$G_{E0}$, $G_{M1}$ and $G_{E2}$ 
form factors \cite{Alexandrou10}. 
Under these conditions, and given that the term from 
the S-state dominates in $G_{E0}$ and $G_{M1}$,
the most important constraint for the D-states
comes from $G_{E2}$ (to which only 
the interference term between S- and D3-state contributes), and
a decisive constraint on the D1 state 
(through $G_{M3}$) is not available yet,
except for the quenched calculation of  Ref.~\cite{Boinepalli09} 
for $Q^2=0.23$ GeV$^2$.
That point is also considered in our fit.

In principle one could assume that the D1 state would 
not be important for the  $\Omega^-$  structure --- we evoke  
that it was checked earlier
that the lattice data for the $\Delta$ baryon can be described 
with very small ($\approx 1\%$) D state admixtures \cite{LatticeD}.
Still, this needs to be investigated, and therefore we check here 
whether the D1 state can be important
to $G_{E0}$ and $G_{M1}$ data and helps to improve the overall
description of the form factor data. 
This is why in this work we leave the D1-state 
mixture free and use the lattice data
(mainly $G_{M1}$) to fix that contribution.
Future lattice QCD simulations  for $G_{M3}$ 
may then confirm or contradict our model.

The only limitation of our calculation
is that the amount of D-states admixture
is assumed to be small, since the calculation of 
the form factor  proceeds by taking 
only into account the first order
terms  in $a$ and $b$ (neglecting transitions between D-states).
That limitation will be quantified at the end,
comparing the result of  
$G_{E0}(0)$ with the 
$\Omega^-$ charge ($-1$), obtained 
when all states are 
include in the electromagnetic transition current.

\subsection{Lattice data}

We use the lattice QCD data from 
Alexandrou \etal \cite{Alexandrou10}
and the $G_{M3}$ datapoint from Ref.~\cite{Boinepalli09}. 
All the lattice QCD simulations 
from Ref.~\cite{Alexandrou10}
are unquenched 
but two different methods are used: 
hybrid action and domain wall fermions (DWF).
The single datapoint from Ref.~\cite{Boinepalli09} is quenched.
The data from Ref.~\cite{Alexandrou10}
goes almost to 4 GeV$^2$ \cite{PrivateC}, but 
only the data for $Q^2 < $ 1.6 GeV$^2$ is relatively precise.
The simulations were performed for one pion mass 
using the hybrid method ($m_\pi=0.353$ GeV) and 
three pion masses with the DWF method 
($m_\pi= 0.297,\, 0.330$ and 0.355 GeV).

The results of $G_{E0}$ and $G_{M1}$  in the DWF case  
show a weak dependence in the pion mass, 
suggesting that meson cloud effects ($K$ and $\eta$) 
may be negligible at least for those form factors 
as expected from a three strange valence quark system.
The hybrid simulation for  $G_{E0}$ and $G_{M1}$ 
shows a systematic deviation 
from the results from DWF (slower falloff) for large $Q^2$ 
with similar pion mass ($m_\pi=0.355$ GeV).


For the hybrid simulation
one has results for $G_{E0}$, $G_{M1}$ and $G_{E2}$ for pion mass value of  $m_\pi= 0.353$ GeV.
As for the DWF simulations, one has data for $G_{M1}$, with $m_\pi=0.355$ GeV;
for $G_{E0}$ and $G_{M1}$ with  $m_\pi=0.330$ GeV;
and finally for  data for $G_{E0}$, $G_{M1}$ 
and also for $G_{E2}$ with $m_\pi=0.297$ GeV.
The difference between the two methods,
particularly to high $Q^2$, 
can be a lattice artifact (cutoff effect)  \cite{Alexandrou10}.
More lattice QCD simulations with smaller 
lattice spacing and larger volumes are necessary 
to clarify the differences between 
the two methods  \cite{Alexandrou10}.

The  $\Omega^-$ form factors can also be 
obtained  from a simulation of the $\Delta^-$ in the SU(3) symmetry limit.
This was done by Boinepalli \etal \cite{Boinepalli09} 
for pion mass $m_\pi= 0.697$ GeV, 
leading to a slightly heavier mass for the $\Omega^-$
(1.732 GeV to be compared with 1.672 GeV of the physical case \cite{PDG}).
This simulation is quenched and the results are 
given only for one value of $Q^2$ 
[$Q^2=0.23$ GeV$^2$]. Importantly,
that work provides the only existing clue 
for the behavior of the octupole magnetic 
moment form factor:
$G_{M3}(0.23\, \mbox{GeV}^2)=1.25\pm7.50$.


\begin{table}[t]
\begin{center}
\begin{tabular}{c c c c c c }
\hline
\hline
$\chi^2(G_{E0})$ & $\chi^2(G_{M1})$ &  $\chi^2(G_{E2})$  &  $\chi^2(G_{M3})$
& $\chi^2(tot)$ 
& $G_{E2}(0)$\\
\hline
5.28            &    2.66   & 0.203 & 2.49 & 2.58 & 0.674 \\
\hline
\hline
\end{tabular}
\end{center}
\caption{Results of the $\chi^2$ in the fit to the lattice data.
The results correspond to the values: 
$a=$0.0341, $b=0.2666$, $\alpha_S=0.1793$, 
$\alpha_{D3}=0.5394$ and $\alpha_{D1}=0.4674$.}
\label{tableChi}
\end{table}

\subsection{Calibrating the model (fit to lattice data)}

Ideally, to extrapolate the $\Omega^-$ form factors 
from lattice QCD one should take the set of 
lattice QCD simulations performed as close as possible 
to the physical limit -- 
the lowest pion mass ($m_\pi= 0.297$ GeV)
considered in the DWF case.
Unfortunately, the available  DWF data for $G_{E2}$ 
 is restricted to 7 data points,
for that mass,  
 below 2 GeV$^2$, 
and there is no data for $Q^2 < 0.4$ GeV$^2$, 
which is a severe limitation 
to extrapolate the results down to $Q^2=0$. 
To obtain a more accurate extrapolation, we took also the hybrid data 
(with 13 datapoints for $G_{E2}$ below 2 GeV$^2$).
To consider simultaneously the hybrid and DWF data increases 
the statistics, which in principle leads 
to more accurate constraint of the model.
However, the differences between the two methods, 
in particular the high accuracy of the $G_{E0}$ data, 
make a good fit difficult, particularly for the
$G_{E0}$ and $G_{M1}$ form factors.
Since the difference between the two methods 
is amplified as $Q^2$ increases (particularly for $Q^2 > 1$ GeV$^2$),
to improve the quality of our fit 
we took  only the $G_{E0}$ and $G_{M1}$ data
below $Q^2 \le 1$ GeV$^2$. 
This is justified since the
lattice simulations have larger numerical error for higher $Q^2$
and also because we are focused in the 
extrapolation for the $Q^2=0$ point,  which leads us to reduce  
the weight of high $Q^2$ data.
This procedure is frequently 
used in lattice QCD studies at low $Q^2$ \cite{Syritsyn10}.
The inclusion of the region $Q^2> 1$ GeV$^2$
in the fit will be possible once a more homogeneous 
set of data with sufficient statistics is provided.
As for $G_{E2}$  we took the data for $Q^2 < 2$ GeV$^2$, 
since the errorbars are more significant and we want to 
keep the statistics as large as possible.

The results for the $\chi^2$ of the fit are presented 
in the Table \ref{tableChi}.
The large $\chi^2$ for the $G_{E0}$ data is 
a consequence of the great precision 
of the data and also of  the different behavior 
of the hybrid and DWF data.
The parameters of the fit are indicated in
the caption of Table \ref{tableChi}.
The mixture of D-states that we obtain is  0.11\% (for D3) 
and 6.6\% (for D1). 
There is therefore a significant mixture of the D1 state. 

In detail, the value for the momentum-range parameter $\alpha_S$
(0.1793)  obtained in this fit (with S and D states) 
suggests that D states improve in fact the 
description of $G_{E0}$ and $G_{M1}$, 
when compared with the predictions from 
Ref.~\cite{Omega},
prior to the simulations of Ref.~\cite{Alexandrou10}
and where $\alpha_S$ was smaller (0.1630).

\begin{figure*}[t]
\centerline{
\mbox{
\includegraphics[width=2.9in]{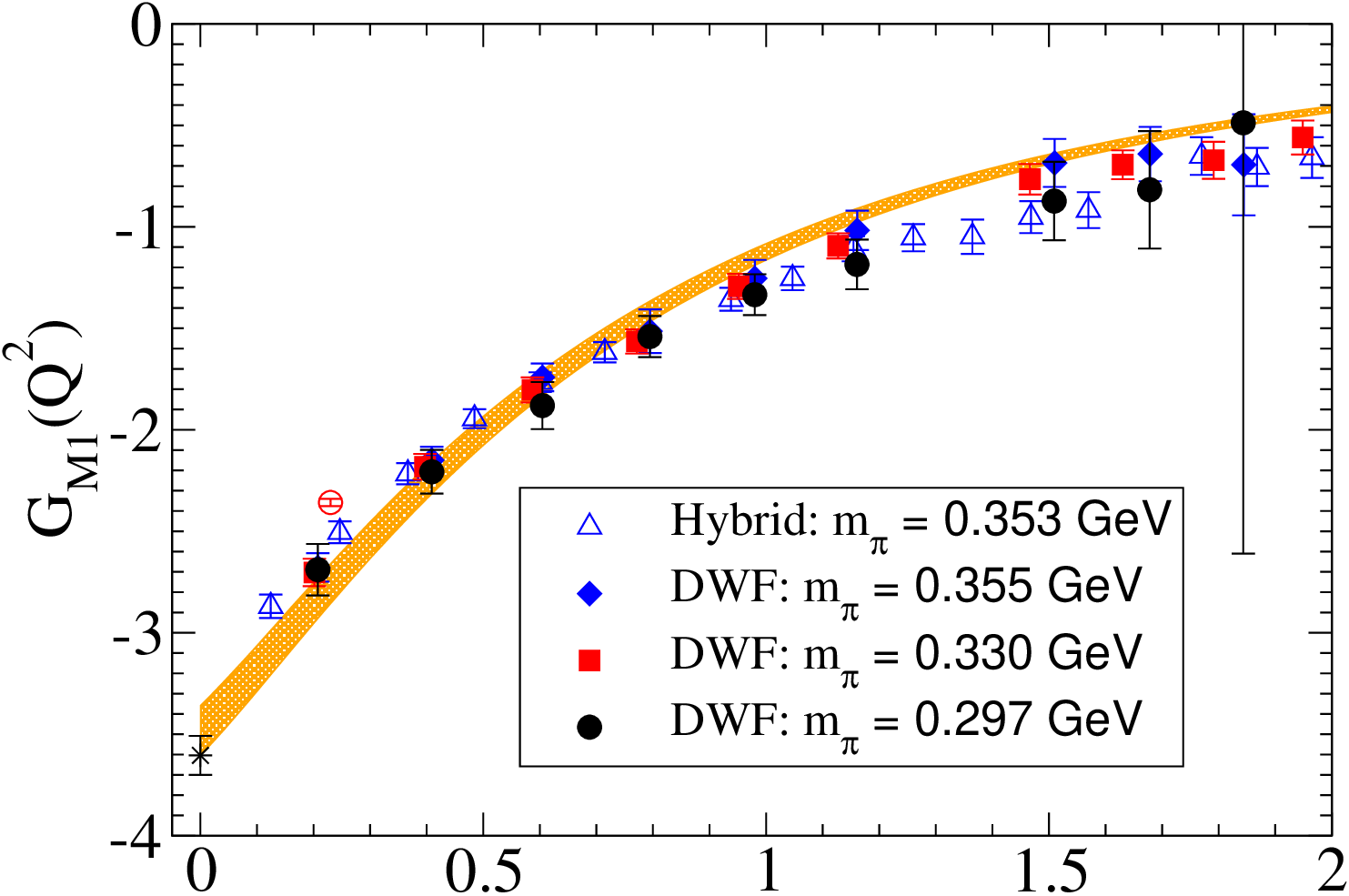} 
\includegraphics[width=2.9in]{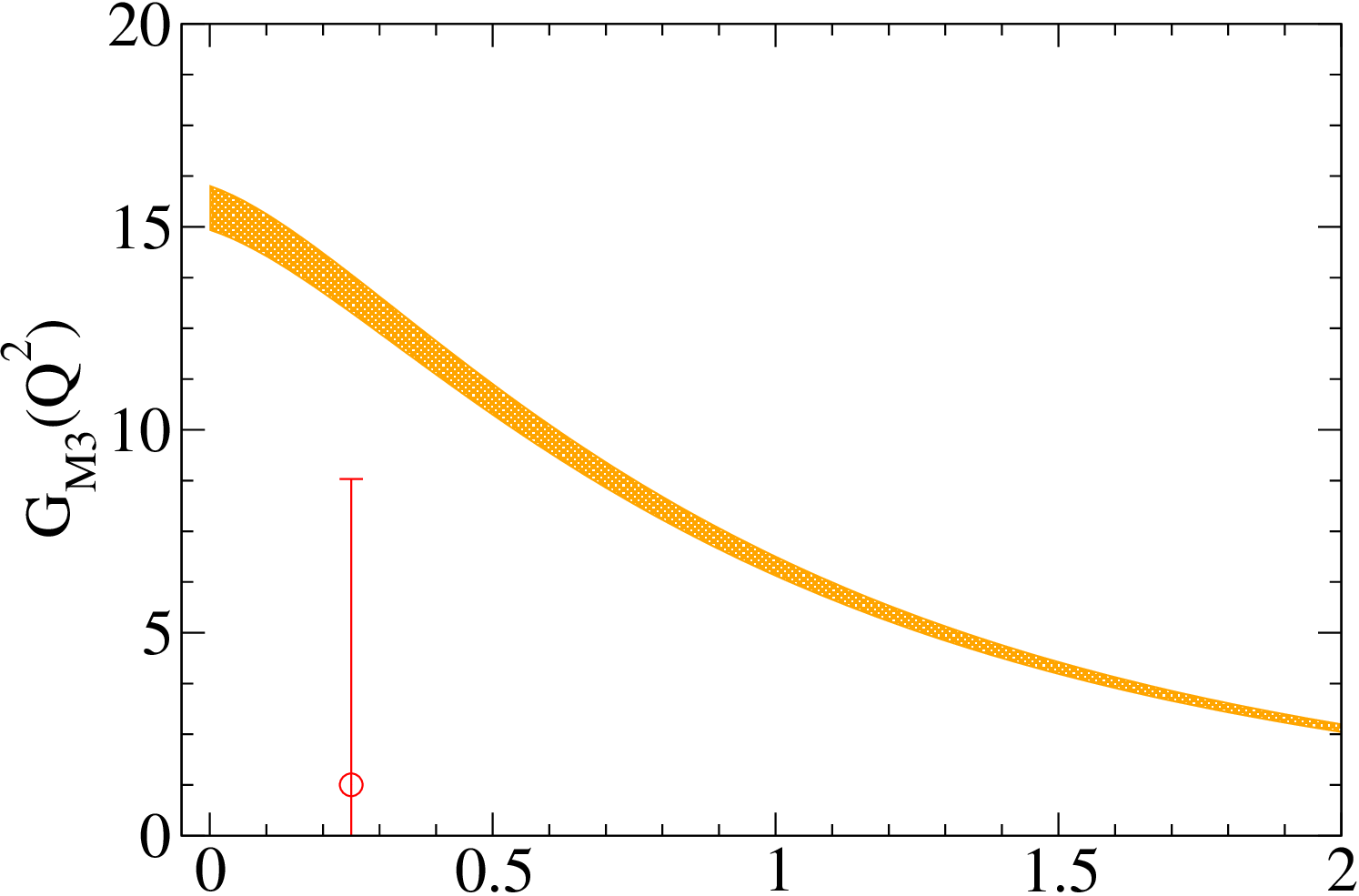}
}}
\vspace{.9cm}
\centerline{
\mbox{
\includegraphics[width=2.9in]{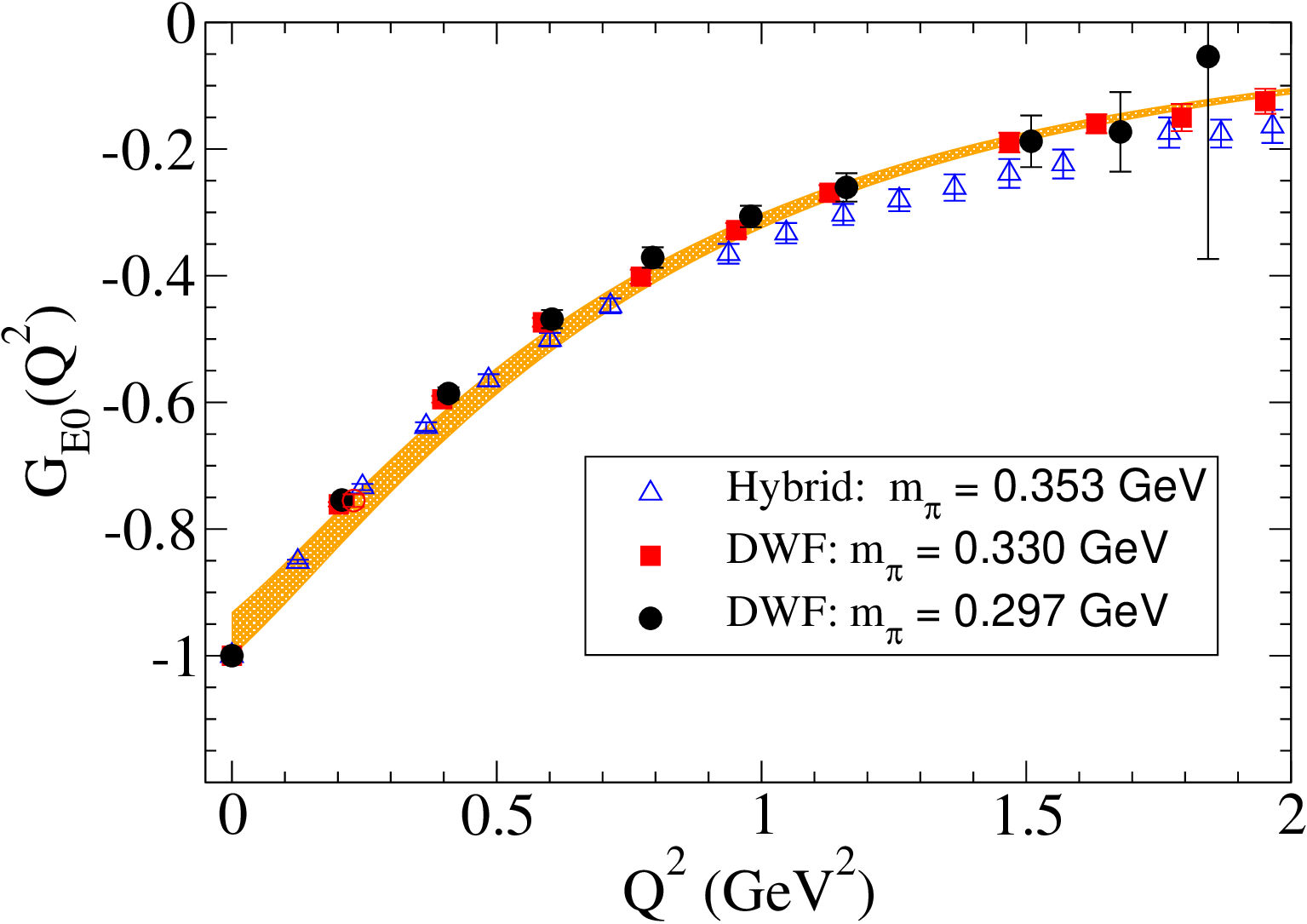} 
\includegraphics[width=2.9in]{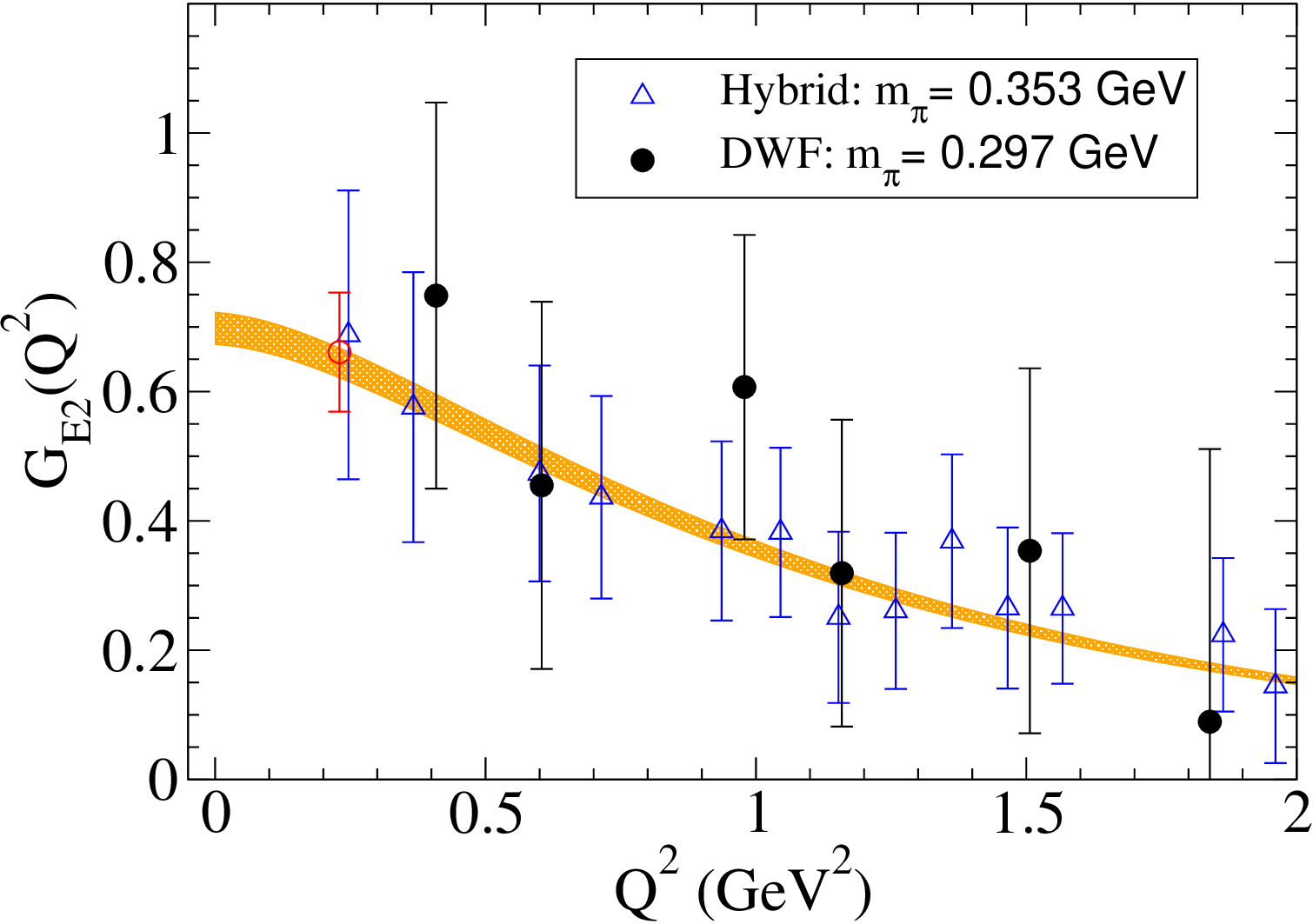}
}}
\caption{\footnotesize{Best fit with a D1 and D3 mixture.
Lattice QCD data from Alexandrou \etal \cite{Alexandrou10}.
For $G_{M1}$ we include also the experimental result 
$G_{M1}(0)= -3.604   \pm 0.096$ from PDG \cite{PDG} 
($\ast$). The open circles represent the 
result for $Q^2=0.23$ GeV$^2$ from
Boinepalli \etal \cite{Boinepalli09}.}}
\label{figFF}
\end{figure*}

The results of our fit to the lattice form factors
are presented in Fig.~\ref{figFF}.
The  upper (lower) results   for $G_{E0}$ and $G_{M1}$ 
($G_{E2}$ and $G_{M3}$) are given by Eqs.~(\ref{eqGE0a})-(\ref{eqGM3a}),
with the factor $N^2=1/(1+ a^2+ b^2)$ defined 
by the normalization of the $\Omega^-$ wavefunction\footnote{Note that  
in the present work, by taking 
$N^2=1/(1+ a^2+ b^2)$, we are not modifying the normalization 
of the $\Omega^-$ wavefunction  relatively to Ref.~\cite{Omega}. 
In previous calculations  
we had $N^2=1$ because  only the S state 
(normalized to 1) was considered. Once the D- states are added, 
$N^2$ is redefined accordingly to the D- state
admixture coefficients to $N^2=1/(1+ a^2+b^2)$.},
given by Eq.~(\ref{eqPsiOm}).
As mentioned in Sect.~\ref{secFormFactors}, those equations are derived in 
first order in the D-state admixture parameters $a$ and $b$,
and differ from the final result because 
the transitions between D-states are neglected.
We can calculate the effect of this approximation
by evaluating the correction to  $G_{E0}(Q^2)$
needed to reproduce exactly the $\Omega^-$ charge at $Q^2=0$.
From Eq.~(\ref{eqGE0a})
the exact result for $G_{E0}(0)$ 
is obtained  by setting $N^2 = 1$,  which then defines 
our lowest estimation of $G_{E0}$.
The band between the upper and the lower result
measures the effect in the final result of the neglected
D to D-state transitions.
The width of the band is small because  
$a$ and $b$ turn out to be small from the fit to the lattice data.
As for the uncertainty in the calculation of the form factors, $G_{M1}$,
$G_{E2}$ and $G_{M3}$ we use the same procedure, 
for consistency.

Our results presented in Fig.~\ref{figFF}, for 
the four form factors, are consistent 
with the overall lattice data within one standard deviation, to
the exception of the form factor $G_{M3}$.
In that case, although we
overestimate the lattice datapoint \cite{Boinepalli09},
our result is still inside an uncertainty interval 
of  1.5 standard deviations.
Note that in our formalism the 
magnitude of $G_{M3}$ is a consequence of the
relatively large D1 state admixture (6.6\%),  a prediction 
to be tested by future and more precise lattice QCD simulations. 
The extracted value of the electric quadrupole 
form factor at $Q^2=0$ is $G_{E2}(0)=0.680\pm0.012$.

\section{Conclusions}
\label{secConclusions}

The $\Omega^-$ is the most stable baryon with spin $3/2$.
Yet, only some of its properties are known.
As a spin 3/2 particle with a long lifetime, it is the first candidate 
for the experimental determination of the electric quadrupole moment,
since the nucleon (spin 1/2) has no quadrupole moment, and the lifetime
of the $\Delta$ is much shorter.
Several experimental methods have been
proposed to measure the still unknown quadrupole moment
$Q_{\Omega^-}$,  which is a signature of distortion and is likely to be
measured in a very near future.
This makes its a-priori prediction so challenging.

As the $\Omega^-$ is essentially 
a three strange valence quark system it is 
possible nowadays to simulate the electromagnetic coupling 
with the $\Omega^-$ in a discrete lattice for 
physical strange quark masses with light sea quarks 
($m_\pi \simeq 300$ MeV) and to determine the $\Omega^-$ 
electromagnetic form factors.
In the absence of
experimental information, lattice QCD provides 
then the more reliable method 
to unveil the $\Omega^-$ electromagnetic structure.

However, to determine the electric quadrupole moment
an extrapolation of the $G_{E2}$ form factor 
down to $Q^2=0$ must be done, and 
some analytical form near the origin must be assumed.
In this work we provide then a method to extract information 
from the lattice QCD data, without assuming any special analytical form, 
as a  dipole, tripole or exponential function.
We start by requiring an overall consistency between 
the prediction of our model for the four form factors 
and the lattice QCD data for the physical $\Omega^-$ mass.
Then, the model extends naturally down to the $Q^2=0$  limit,
inferring the behavior 
of the $\Omega^-$ system in the region not 
covered by the lattice data.
As a bonus relatively to an ad-hoc parameterization,
we obtain even information about the microscopic structure
such as the  admixture percentage of each D- state
and the momentum distribution of the wavefunction.

The procedure assumes that meson cloud dressing 
is not significant in the $\Omega^-$ system.
Although the present lattice QCD results
cannot rule out that sea quark dressing 
(meson cloud) may be important, they suggest that dependence
in the light quark (pion mass) is small at least 
for the charge and magnetic dipole form factors.
This topic will be investigated in the future.

Our final results imply an unexpected large D1 state mixture (6.6\%).
The confirmation or disproval of this result will 
be possible once lattice data 
for $G_{M3}$ became available.
More precise data for $G_{M1}$ ($G_{E0}$ is already very precise), 
particularly at high $Q^2$, where the weight of 
the D states is larger, 
will also be useful for an even better estimate 
of the both D-state admixtures. 

Our final result for the electric quadrupole moment of the 
$\Omega^-$ is $Q_{\Omega^-}=(0.96\pm0.02)\times 10^{-2}e$fm$^2$.  
In the literature the existing results for 
$Q_{\Omega^-}$ correspond to the interval
$(0.4-4.0) \times 10^{-2}e$fm$^2$
\cite{Gershtein81,Richard82,Isgur82,Leonard90,Krivoruchenko91,Butler94,Kroll94,Wagner00,Buchmann02,Buchmann07,Geng09,Nicmorus10}, and our result has a magnitude consistent with this range.
The value extracted directly from lattice QCD data, assuming an exponential dependence 
\cite{Alexandrou10},  corresponds to $(1.18\pm0.12) \times 10^{-2}e$fm$^2$.
Note that our result satisfies
simultaneously the constraints of the three 
form factors ($G_{E0}$, $G_{M1}$ and $G_{E2}$),  
and therefore can be given with  a smaller band of uncertainty.

The numerical value of $Q_{\Omega^-}/e_{\Omega^-}$ 
($e_{\Omega^-}=-1$), because it is a positive number,
can be interpreted in a nonrelativistic formalism 
as the charge distribution being
extended and flattened along the equatorial region, 
as it was also predicted for the $\Delta^+$  baryon
\cite{DeltaDFF}.
Note however that this interpretation 
has been questioned,  and some authors suggest
a different concept of deformation, 
based on the transverse  electric quadrupole moment 
in the infinite momentum frame
\cite{Alexandrou09b,Alexandrou09,Deformation}.  

Comparing the $\Omega^-$ with the $\Delta^+$ baryon,
in the covariant spectator quark model 
it is interesting to notice that the difference 
in the D3 admixture, 0.72\%
for the $\Delta$ and 0.11\% for the $\Omega^-$ 
does not correspond to a reduction of 
$G_{E2}(0)$ in the same proportion.
This shows that the momentum distribution in the overlap integral 
between S and D3 states is very different in both systems 
and it has to be taken 
into consideration.

An intrinsic limitation of our calculation
is the determination of the form factors 
only in first order of the admixture coefficients.
The exact calculation of the form factors 
including the contributions from transitions between 
D states is in progress \cite{FormFactorsDD}.
Nevertheless, the effect of these transitions
can be estimated approximately by 
the correction implied by the deviation of the
normalization constant $N$ from 1. 
In the present case this correction is about 3.4\%.

Finally, in the future, a precise calculation of the 
magnetic octupole form factor of the $\Omega^-$
using lattice QCD will be very useful 
for a better understanding of the $\Omega^-$ 
electromagnetic structure.
The  magnetic octupole moment 
${\cal O}_{\Omega^-}= G_{M3}(0)\sfrac{e}{2M_\Omega^3}$,
was estimated already  in 
Refs.~\cite{Buchmann08,Aliev09,Geng09}.
Our result corresponds to 
$(1.27\pm0.04)\times 10^{-2}e$fm$^3$.
Although the experimental determination of ${\cal O}_{\Omega^-}$
may not be possible in practice, an evaluation
based on QCD in a regime where meson 
excitations are expected to be small,
may help to select models.
In particular for our model, it can 
allow a more accurate determination 
of the  wavefunction  structure, 
which carry information 
on the shape of the electromagnetic distributions
inside that baryon, and also the 
the relative contribution of the 
D-wave components in the wavefunction of the $\Omega^-$.

\vspace{0.2cm}
\noindent
{\bf Acknowledgments:}

The authors thanks Constantia Alexandrou and collaborators, 
in particular to Yiannis Proestos and 
Tomasz Korzec and for providing 
the data 
from Ref.~\cite{Alexandrou10} 
and for the detailed explanations. 
The authors thanks also Dru Renner, 
Kazuo Tsushima for helpful discussions,
Franz Gross and Jefferson Lab Theory Center 
for the hospitality during the period of conclusion of this paper.
G.~R.~was supported by the Portuguese Funda\c{c}\~ao para
a Ci\^encia e Tecnologia (FCT) under the Grant
No.~SFRH/BPD/26886/2006.
This work has been supported in part by the European Union
(HadronPhysics2 project ``Study of strongly interacting matter'').


\begin{thebibliography}{00}





\bibitem{DeltaDFF}
  G.~Ramalho, M.~T.~Pe\~na and F.~Gross,
  Phys.\ Lett.\  B {\bf 678}, 355 (2009)
  [arXiv:0902.4212 [hep-ph]].

\bibitem{DeltaDFF2}
  G.~Ramalho, M.~T.~Pe\~na and F.~Gross,
  Phys.\ Rev.\  D {\bf 81}, 113011 (2010)
  [arXiv:1002.4170 [hep-ph]].








\bibitem{Gershtein81}
  S.~S.~Gershtein and Yu.~M.~Zinovev,
  Sov.\ J.\ Nucl.\ Phys.\  {\bf 33}, 772 (1981)
  [Yad.\ Fiz.\  {\bf 33}, 1442 (1981)].

\bibitem{Richard82}
  J.~M.~Richard,
  Z.\ Phys.\  C {\bf 12}, 369 (1982).


\bibitem{Isgur82}
  N.~Isgur, G.~Karl and R.~Koniuk,
  Phys.\ Rev.\  D {\bf 25}, 2394 (1982).




\bibitem{Leonard90}
  W.~J.~Leonard and W.~J.~Gerace,
  Phys.\ Rev.\  D {\bf 41}, 924 (1990).


\bibitem{Krivoruchenko91}
  M.~I.~Krivoruchenko and M.~M.~Giannini,
  Phys.\ Rev.\  D {\bf 43}, 3763 (1991).





\bibitem{Wagner00}
  G.~Wagner, A.~J.~Buchmann and A.~Faessler,
  J.\ Phys.\ G {\bf 26}, 267 (2000).




\bibitem{Dahiya10}
  H.~Dahiya and N.~Sharma,
  arXiv:1009.1950 [hep-ph].






\bibitem{Butler94}
  M.~N.~Butler, M.~J.~Savage and R.~P.~Springer,
  Phys.\ Rev.\  D {\bf 49}, 3459 (1994)
  [arXiv:hep-ph/9308317].


\bibitem{Geng09}
  L.~S.~Geng, J.~Martin Camalich and M.~J.~Vicente Vacas,
  Phys.\ Rev.\  D {\bf 80}, 034027 (2009)
  [arXiv:0907.0631 [hep-ph]].







\bibitem{Buchmann02}
   A.~J.~Buchmann and E.~M.~Henley,
   Phys.\ Rev.\  D {\bf 65}, 073017 (2002).



\bibitem{Buchmann03}
  A.~J.~Buchmann and R.~F.~Lebed,
  Phys.\ Rev.\  D {\bf 67}, 016002 (2003)
  [arXiv:hep-ph/0207358].







\bibitem{Buchmann07}
  A.~J.~Buchmann,
  AIP Conf.\ Proc.\  {\bf 904}, 110 (2007)
  [arXiv:0712.4270 [hep-ph]].






\bibitem{Kroll94}       
  J.~Kroll and B.~Schwesinger,
  Phys.\ Lett.\  B {\bf 334}, 287 (1994)
  [arXiv:hep-ph/9405267].


\bibitem{Oh95}    
  Y.~Oh,
  Mod.\ Phys.\ Lett.\  A {\bf 10}, 1027 (1995)
  [arXiv:hep-ph/9506308].



\bibitem{Aliev09}
  T.~M.~Aliev, K.~Azizi and M.~Savci,
  Phys.\ Lett.\  B {\bf 681}, 240 (2009)
  [arXiv:0904.2485 [hep-ph]].



\bibitem{Nicmorus10}
  D.~Nicmorus, G.~Eichmann and R.~Alkofer,
  arXiv:1008.3184 [hep-ph].






\bibitem{PDG}
  C.~Amsler {\it et al.}  [Particle Data Group],
  Phys.\ Lett.\  B {\bf 667}, 1 (2008).





\bibitem{Omega}
  G.~Ramalho, K.~Tsushima and F.~Gross,
  Phys.\ Rev.\  D {\bf 80}, 033004 (2009)
  [arXiv:0907.1060 [hep-ph]].








\bibitem{Diehl91}
  H.~T.~Diehl {\it et al.}
  Phys.\ Rev.\ Lett.\  {\bf 67}, 804 (1991).



\bibitem{Wallace95}
  N.~B.~Wallace {\it et al.},
  Phys.\ Rev.\ Lett.\  {\bf 74}, 3732 (1995).








 
\bibitem{Sternheimer73}    
  R.~M.~Sternheimer and M.~Goldhaber,
  Phys.\ Rev.\  A {\bf 8}, 2207 (1973).



\bibitem{Baryshevsky93}
  V.~G.~Baryshevsky and A.~G.~Shechtman,
  Nucl.\ Instrum.\ Meth.\  B {\bf 83} 250 (1993).



\bibitem{Karl07}
   G. Karl, 
   {\it 
   Proceedings of 
   IVth International Conference on Quarks and Nuclear Physics,
   Part 3}, 155 (2007)





\bibitem{Krivoruchenko08}
  M.~I.~Krivoruchenko and A.~Faessler,
  Nucl.\ Phys.\  A {\bf 803}, 173 (2008)
  [arXiv:0707.3050 [nucl-th]].













\bibitem{Bernard82}
  C.~W.~Bernard, T.~Draper, K.~Olynyk and M.~Rushton,
  Phys.\ Rev.\ Lett.\  {\bf 49}, 1076 (1982).


\bibitem{Leinweber92}     
  D.~B.~Leinweber, T.~Draper and R.~M.~Woloshyn,
  Phys.\ Rev.\  D {\bf 46}, 3067 (1992)
  [arXiv:hep-lat/9208025].



\bibitem{Aubin}
  C.~Aubin, K.~Orginos, V.~Pascalutsa and M.~Vanderhaeghen,
  Phys.\ Rev.\ D {\bf 79}, 051502(R) (2009)
  arXiv:0811.2440 [hep-lat].




\bibitem{Alexandrou10}
  C.~Alexandrou, T.~Korzec, G.~Koutsou, J.~W.~Negele and Y.~Proestos,
  Phys.\ Rev.\  D {\bf 82}, 034504 (2010)
  [arXiv:1006.0558 [hep-lat]].


\bibitem{Boinepalli09}
  S.~Boinepalli, D.~B.~Leinweber, P.~J.~Moran, 
  A.~G.~Williams, J.~M.~Zanotti and J.~B.~Zhang,
  Phys.\ Rev.\  D {\bf 80}, 054505 (2009)
  [arXiv:0902.4046 [hep-lat]].







\bibitem{Nucleon}
  F.~Gross, G.~Ramalho and M.~T.~Pe\~na,
  Phys.\ Rev.\  C {\bf 77}, 015202 (2008)
  [arXiv:nucl-th/0606029].



\bibitem{FixedAxis}
  F.~Gross, G.~Ramalho and M.~T.~Pe\~na,
  Phys.\ Rev.\  C {\bf 77}, 035203 (2008).



\bibitem{Octet}
  F.~Gross, G.~Ramalho and K.~Tsushima,
  Phys.\ Lett.\  B {\bf 690}, 183 (2010)
  [arXiv:0910.2171 [hep-ph]].






\bibitem{Lattice}
  G.~Ramalho and M.~T.~Pe\~na,
  J.\ Phys.\ G {\bf 36}, 115011 (2009)
  [arXiv:0812.0187 [hep-ph]].



\bibitem{DeltaFF0}
  G.~Ramalho and M.~T.~Pe\~na,
  J.~Phys.~G {\bf 36}, 085004 (2009)
  arXiv:0807.2922 [hep-ph].


\bibitem{Delta1600}
  G.~Ramalho and K.~Tsushima,
  Phys.\ Rev.\  D {\bf 82}, 073007 (2010)
  [arXiv:1008.3822 [hep-ph]].




\bibitem{NDelta}
  G.~Ramalho, M.~T.~Pe\~na and F.~Gross,
  Eur.\ Phys.\ J.\  A {\bf 36}, 329 (2008)
  [arXiv:0803.3034 [hep-ph]].

\bibitem{NDeltaD}
  G.~Ramalho, M.~T.~Pe\~na and F.~Gross,
  Phys.\ Rev.\  D {\bf 78}, 114017 (2008)
  [arXiv:0810.4126 [hep-ph]].


\bibitem{LatticeD}
  G.~Ramalho and M.~T.~Pe\~na,
  Phys.\ Rev.\ D {\bf 80}, 013008 (2009)
  arXiv:0901.4310 [hep-ph].





\bibitem{Roper}
  G.~Ramalho and K.~Tsushima,
  Phys.\ Rev.\  D {\bf 81}, 074020 (2010)
  [arXiv:1002.3386 [hep-ph]].



\bibitem{ExclusiveR}
  G.~Ramalho, F.~Gross, M.~T.~Pe\~na and K.~Tsushima,
  arXiv:1008.0371 [hep-ph].






\bibitem{PrivateC}
  Y.~Proestos and  T.~Korzec,
  private comunication.




\bibitem{Syritsyn10}
  S.~N.~Syritsyn {\it et al.},
  Phys.\ Rev.\  D {\bf 81}, 034507 (2010)
  [arXiv:0907.4194 [hep-lat]].




\bibitem{Alexandrou09b}
  C.~Alexandrou {\it et al.},
  Phys.\ Rev.\  D {\bf 79}, 014507 (2009)
  [arXiv:0810.3976 [hep-lat]].

\bibitem{Alexandrou09}
  C.~Alexandrou {\it et al.},
  Nucl.\ Phys.\  A {\bf 825}, 115 (2009)
  [arXiv:0901.3457 [hep-ph]].


\bibitem{Deformation}
  G.~Ramalho, M.~T.~Pe\~na and A.~Stadler,
  work in preparation




\bibitem{FormFactorsDD}
  G.~Ramalho and M.~T.~Pe\~na, 
  work in preparation.



\bibitem{Buchmann08}
  A.~J.~Buchmann and E.~M.~Henley,
  Eur.\ Phys.\ J.\  A {\bf 35}, 267 (2008)
  [arXiv:0808.1165 [hep-ph]].



\end{thebibliography}
\end{document}